\documentclass[twocolumn]{aastex62}
\shorttitle{Relativistic Astronomy}
\shortauthors{Zhang \& Li}
\begin{document}


\title{Relativistic Astronomy}

\correspondingauthor{Bing Zhang}
\email{zhang@physics.unlv.edu}

\author{Bing Zhang}
\affiliation{Department of Physics and Astronomy, University of Nevada Las Vegas, NV 89154}
%\nocollaboration

\author{Kunyang Li}
\affiliation{Department of Physics, 837 State St NW, Georgia Institute of Technology, Atlanta, GA 30332}
\affiliation{Center for Relativistic Astrophysics, Georgia Institute of Technology, Atlanta, GA 30332}
%\nocollaboration


%\author{Bing Zhang$^{1}$, Kunyang Li$^{2,3}$}
%\affil{1. Department of Physics and Astronomy, University of Nevada Las Vegas, NV 89154, USA; zhang@physics.unlv.edu}
%\affil{2. Department of Astronomy, School of Physics, Peking University, Beijing, 100871, P. R. China}
%\affil{3. Kavli Institute for Astronomy and Astrophysics, Peking University, Beijing, 100871, P. R. China}
%\affil{2. Department of Physics, 837 State St NW, Georgia Institute of Technology, Atlanta, GA 30332, USA}
%\affil{3. Center for Relativistic Astrophysics, Georgia Institute of Technology, Atlanta, GA 30332, USA}



\begin{abstract}
The ``Breakthrough Starshot'' aims at sending near-speed-of-light cameras to nearby stellar systems in the future. Due to the relativistic effects, a trans-relativistic camera naturally serves as a spectrograph, a lens, and a wide-field camera. We demonstrate this through a simulation of the optical-band image of the nearby galaxy M51 in the rest frame of the trans-relativistic camera. We suggest that observing celestial objects using a trans-relativistic camera may allow one to study the astronomical objects in a special way, and to perform unique tests on the principles of special relativity. We outline several examples that trans-relativistic cameras may make important contributions to astrophysics and suggest that the Breakthrough Starshot cameras may be launched in any direction to serve as a unique astronomical observatory.
\end{abstract}





\section{Introduction}

The Breakthrough Initiatives (https://breakthroughinitiatives.org/) are a program of scientific and technological exploration, probing some big questions of life in the universe. Among them is the ``Breakthrough Starshot'' program (https://breakthroughinitiatives.org/Challenges/3/), which aims at proving the concept of developing unmanned space flight (probe) at a good fraction of the speed of light, $c$. Such a probe is designated to reach nearby stellar systems such as Alpha Centauri within decades, allowing humankind to explore extra-solar systems for the first time. The first prototype ``Sprites'' of 3.5 cm$\times$3.5 cm chips weighing just 4 grams each, which are the precursors to eventual ``starChip'' probes, have been recently launched to a low-earth orbit (https://breakthroughinitiatives.org/News/12). Here we point out that due to the relativistic effects, a trans-relativistic camera naturally serves as a spectrograph, a lens and a wide-field camera while traveling in space, allowing humankind to study the astrophysical objects in a unique manner and to conduct tests on special relativity. 
%Launching trans-relativistic cameras would mark the beginning of ``relativistic astronomy''.

\section{Relativistic Astronomy}

\subsection{Relativistic effects}
When a camera travels in space with a speed close to $c$, some interesting relativistic effects would occur. For example, \cite{christian17} suggested that an interferometer moving at a relativistic speed offers a sensitive probe of acceleration making use of the temporal Terrell effect \citep{terrell59,penrose59}.

Here we focus on the observational distortions of emission from distant astronomical objects.  In the co-moving frame of the probe, all astronomical objects undergo unique Doppler boost (Doppler factor ${\cal D} > 1$) or de-boost (${\cal D} <1$) depending on the Lorentz factor of the probe and the angular between the object with respect to the direction of probe motion. 
For the problem involving a flying probe, one can define two rest frames\footnote{Notice that the definition of $K$ and $K'$ is opposite to the convention when studying relativistically-moving astrophysical objects such as gamma-ray bursts (GRBs). In those studies, the laboratory frame is defined as the frame in which the observer is not moving but the object is moving, which is equivalent to Frame $K'$ defined here. Conversely, the co-moving frame in those studies is defined as the rest frame where the object is not moving, which is equivalent to Frame $K$ defined here.
}: the Earth rest frame or the laboratory frame (which is also the rest frame of astronomical objects), $K$, and the probe co-moving frame, $K'$. Let us define the Lorentz factor of the probe as $\Gamma = 1/\sqrt{1-\beta^2}$, where $\beta = v/c$ is the normalized speed of the probe. In Frame $K'$, all the astronomical objects move with the same Lorentz factor $\Gamma$, but with different angles with respect to the opposite direction of the probe motion. The Doppler factor of the source is defined as \citep{rybicki79}
\begin{equation}
 %\begin{eqnarray}
 {\cal D}  \equiv 
 \frac{1}
 {\Gamma(1-\beta\cos\theta')}
 %^{-1} 
 \equiv \Gamma (1+\beta\cos\theta), 
 \label{eq:Doppler}
\end{equation}
%\end{eqnarray}
where the angle between the object moving direction and the line of sight in two different frames are related through\begin{equation}
 \cos\theta' = \frac{\cos\theta+\beta}{1+\beta\cos\theta}.
 \label{eq:theta}
\end{equation}

Some characteristic angles and Doppler factors are
\begin{itemize}
 \item $\theta' = 0$ \& $\theta=0$: ${\cal D} = {\cal D}_{\rm max} \equiv (1+\beta)\Gamma = \sqrt{\frac{1+\beta}{1-\beta}}$;
 % \simeq 2\Gamma$ (last approximation applies when $\beta \lesssim 1$);
 \item $\theta' = \cos^{-1} \beta $ \& $\theta = \pi/2$: ${\cal D} = \Gamma$;
 \item $\theta'= \theta'_c \equiv \cos^{-1} (\frac{1}{\beta} - \frac{1}{\beta\Gamma})$ \& $\theta= \theta_c \equiv \cos^{-1} (\frac{1}{\beta\Gamma} - \frac{1}{ \beta})$: ${\cal D} = 1$;
 \item $\theta'= \pi/2$ \& $\theta = \cos^{-1}(-\beta)$: ${\cal D} = 1/\Gamma$;
 \item $\theta' = \pi$ \& $\theta = \pi$: ${\cal D} = \frac{1}{(1+\beta)\Gamma}  = \sqrt{\frac{1-\beta}{1+\beta}}$;
% \item When $\theta' \ll 1$ and $\Gamma \gg 1$: ${\cal D} \simeq \frac{2\Gamma}{1+\Gamma^2{\theta'}^2}$; at $\theta' = 1/\Gamma$, ${\cal D} \simeq \Gamma$;
\end{itemize}

The Doppler factor connects the quantities of the source rest frame (frame $K$ in our convention) and those in the probe co-moving frame (frame $K'$ in our convention). In particular, for the camera on board the probe all the emission is blueshifted (redshifted) for ${\cal D} > 1$ (${\cal D} < 1$), i.e.
 \begin{equation}
 \nu'  =  {\cal D} \nu,  \label{eq:nu-D}
 \end{equation}
 
For an isotropic, point source, the specific flux and flux transformations are (Appendix A) % \citep{rybicki79,begelman84,zhang17} 
\begin{eqnarray}
 F'_{\nu'} (\nu') & = & {\cal D}^3 F_\nu (\nu), \nonumber \\ 
 F' (\nu') & = & {\cal D}^4 F (\nu). 
\label{eq:Fnu'-D}
%\nonumber
\end{eqnarray}
For an extended source, the specific flux and flux transformations for an emission pixel reads (Appendix A) %\citep{rybicki79,begelman84,zhang17} (Appendix A)
\begin{eqnarray}
 F'_{\nu'} (\nu') & = & {\cal D} F_\nu (\nu), \nonumber \\ 
 F' (\nu') & = & {\cal D}^2 F (\nu). 
\label{eq:Fnu'-extended-D}
%\nonumber
\end{eqnarray}

These salient relativistic effects open a unique opportunity to study the universe and to test the principle of special relativity, which may be discussed under the broad umbrella of ``relativistic astronomy''.

\subsection{Universe seen from the probe's frame $K'$}
%\begin{itemize}
% \item 
For a wide-field camera moving with a Lorentz factor $\Gamma$, all the objects in the field of view undergo relativistic distortions, including shift of position (Eq.(\ref{eq:theta})), shift of frequency (Eq.(\ref{eq:nu-D})), and change of specific flux and flux (Eqs. (\ref{eq:Fnu'-D}) and (\ref{eq:Fnu'-extended-D})). The degree of distortion, characterized by the Doppler factor ${\cal D}$ (Eq.(\ref{eq:Doppler})), solely depends on the angle $\theta$ with respect to the direction of motion for a constant $\Gamma$. In general, an extended object is bluer and more compact as observed in Frame $K'$. Since astronomers already have a detailed view in Frame $K$, a measurement of the differences in the observed properties between Frame $K$ and $K'$ offers valuable information about the astronomical sources. 

% \item 
For a probe with a constant velocity, based on the position shifts of three point sources (background stars)
 close to the direction of motion (where the sources are squeezed), 
 one can uniquely determine the direction of motion and the Lorentz factor $\Gamma$ (or dimensionless velocity $\beta$) of the probe (Appendix B).

% \item 
Once the direction of motion and $\Gamma$ are determined, one can calculate the Doppler factor of all the celestial objects. Given the same observing frequency of the camera, the intrinsic frequencies of different astronomical objects are different (corrected by the respective ${\cal D}$ of the source). For a direction with $\theta < \theta_c$, one has ${\cal D} > 1$, so that the intrinsic frequency of the source the camera records is redder than the observed frequency. Conversely, for $\theta > \theta_c$ (${\cal D} < 1$), the intrinsic frequency of the source the camera records is bluer than the observed frequency. As a result, for a camera in a particular band (e.g. R band), one may study other frequencies (e.g. infrared (IR) or ultraviolet (UV)) of different sources without the need of using technically challenging IR/UV cameras. In a sense, a relativistically moving camera is a natural spectrograph.

% \item 
In the ${\cal D} > 1$ regime, the fluxes of the sources are enhanced. So a relativistically moving camera is also a natural lens. So in practice, astronomical objects are better studied in the ${\cal D} > 1$ regime, where the source is studied in an intrinsically redder band.

% \item 
If a camera continuously observes an astronomical source as the camera is accelerated, it would record emission of the source in a span of frequencies, so that one can obtain a detailed spectrum of the source in the frequency range between $\nu_{\rm camera}$ and $\nu_{\rm camera} / {\cal D}_{\rm max}$. The higher the achievable ${\cal D}_{\rm max}$, the wider the spectrograph is. For different frequencies, one should properly correct for the respective flux Doppler boosting factors to get the intrinsic flux at those frequencies in order to retrieve the intrinsic spectrum of the source. 

% \item 
The camera itself may be designed to have a grism spectrograph. In this case, after determining the direction of motion and probe Lorentz factor using photometry observations, one may turn on the grism mode to capture the fine spectra of the sources in a different (redder) spectral regime. 

% \item 
Due to the light aberration effect, the objects in the moving direction are more packed. The entire hemisphere in frame $K$ is combed into a cone defined by the angle $\theta=\cos^{-1}\beta$. Given a same field of view, the camera can observe more objects. This effectively increases the field of view of the camera.

\begin{figure}   %%%%% Fig. of M13  simulation
%    \centering
%        \begin{tabular}{@{}c@{}}
\plotone{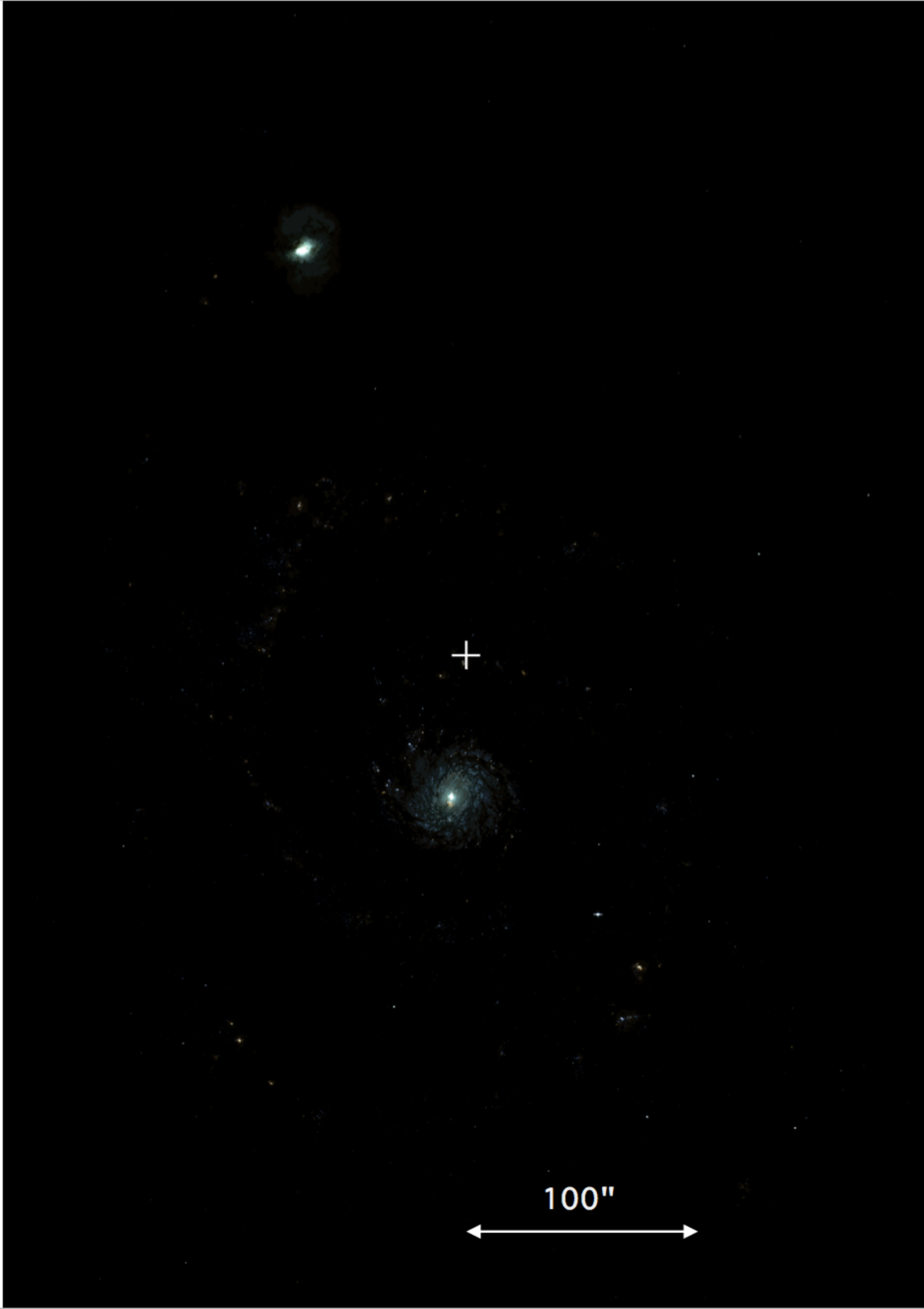}
\plotone{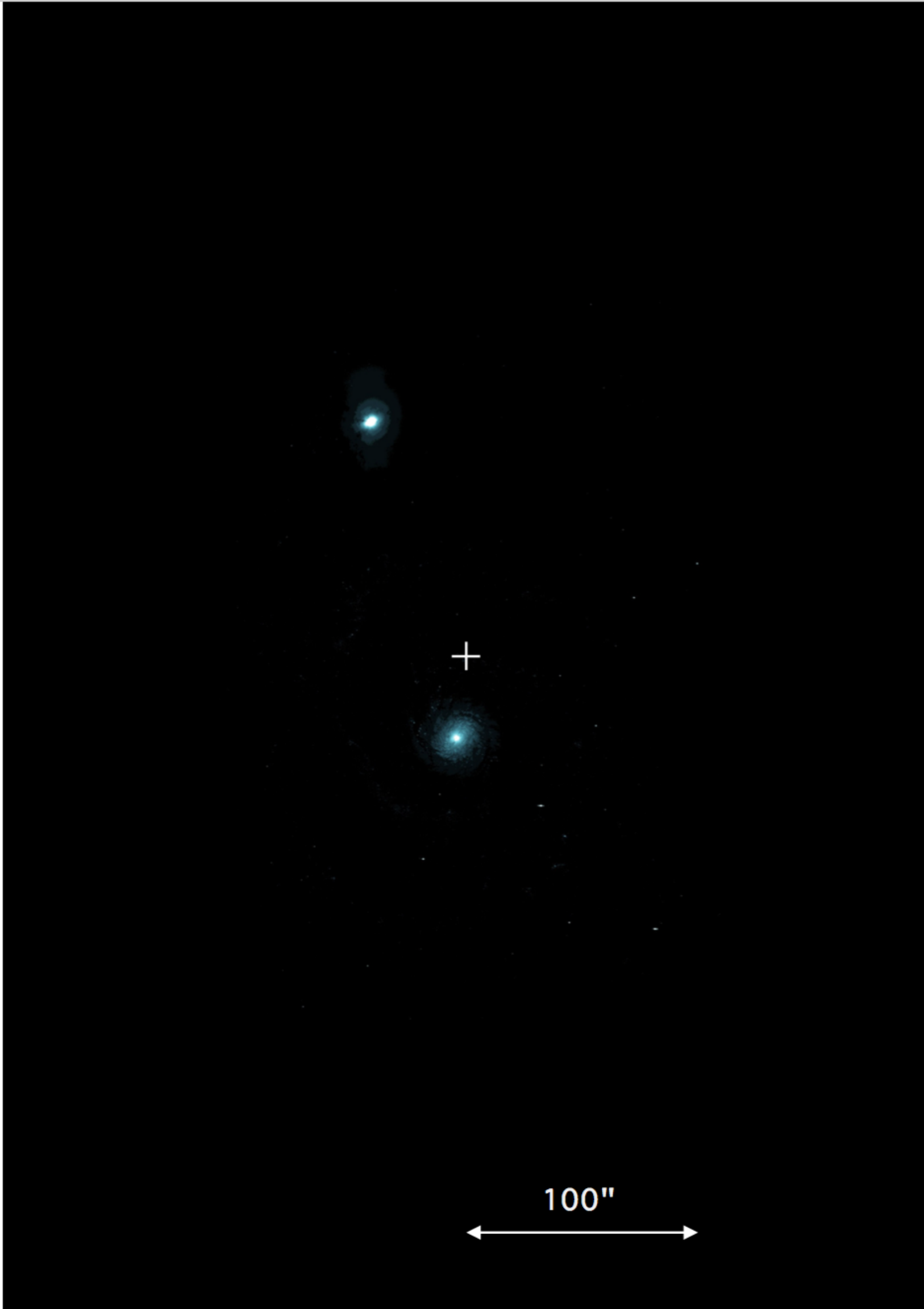}
%        \includegraphics[width=0.4\textwidth]{f1a.pdf}  %{f1a.eps} %{M51_K.png}  
%        \includegraphics[width=0.4\textwidth]{f1b.pdf}  %{f1b.eps} %{M51_Kprime.png}
%        \end{tabular}
    \caption{A comparison of the image of a nearby galaxy M51 in Frames $K$ and $K'$. 
{\it Upper}: False color HST image of M51 in three bands (F435W, F555W, and F814W) observed in the rest frame of Earth, i.e. Frame $K$. 
%Bottom: the same as the top image but zoomed in on the center of the galaxy. 
%The hue values of F435W, F555W, and F814W images are set to 42, 159, and 232. Top: 
{\it Lower}: Simulated false color image of M51 
%in three bands (F435W, F555W, and F658N) 
as observed in the probe's co-moving frame $K'$. The dimensionless velocity is set to $\beta = 0.5$ ($\Gamma \simeq 1.7321$), and the pixel scale of the image is 0.2 arcsec/pixel. The direction of motion is set to the $(1075,1525)$ pixel in the image (cross in both images).} \label{fig:M51}
\label{fig1}
\end{figure}

In order to visually show how the astronomical objects look differently in Frame $K'$, we carry out some simulations based on true observational images. 
%We use  \software{Python, https://www.python.org} and \software{matplotlib, https://matplotlib.org} in generating Figure \ref{fig1}. 
Figure \ref{fig1} upper panel shows an observed image of the nearby galaxy M51. We use the HST images in three filters: F435W, F555W, and F814W\footnote{https://archive.stsci.edu/prepds/m51/datalist.html}. The filter information is from SVO filter profile service website\footnote{http://ivoa.net/documents/Notes/SVOFPS/index.html}. 
In our simulation, we adopt $\beta = 0.5$ ($\Gamma \simeq 1.7321$). 
The simulation code loops through all pixels of the input image, and use the counts of a pixel in different bands to generate a simple spectrum for that particular pixel. Then the spectrum undergoes the relativistic transformation according to Eqs. (\ref{eq:nu-D}) and (\ref{eq:Fnu'-extended-D}). Integration of the product of transformed spectrum and the response functions of the filters give rise to the new counts in different bands of that pixel in frame $K'$. For each pixel, the polar coordinate angles $(\theta, \phi)$ are calculated relative to the direction of motion, which is set to $(1075,1525)$ pixel in the image. Using Eq. (\ref{eq:theta}) the observed polar angle from the moving direction in the probe frame ($\theta'$) can be calculated. With the image scale $(0.2 arcsec/pixel)$, we then obtain the new position of the pixel in the simulated image. In order to simulate the color scheme of the image, we use an image editor named 
%\software{GIMP,https://www.gimp.org}
GIMP (GNU Image Manipulation Program) which enables the simulation with an image in more than three bands. The color of each layer of image is represented by a hue value in GIMP which can be set manually by the user. However, the hue value of images in the same band should be kept the same to show the true color difference between the field of view in frame $K$ and $K'$ due to the relativistic effects.
The hue values of F435W, F555W, and F658N images after relativistic transformation are set to 42, 159, and 232, same as the hue values in the original image. Due to the relativistic effects, the spectrum of M51 is shifted towards blue, meaning that the IR band spectrum is shifted into the optical band. In our simulation, we have used the M51 image in the F814W filter as the IR data, and shifted it to the $H_{\alpha}$ (F658N) filter range in the simulated image. 

The simulated image is presented in the lower panel of Figure \ref{fig1}. One can see that the simulated image is indeed bluer, brighter, and more compact (i.e. each extended source becomes smaller, and the two extended sources have a smaller spatial separation).

\subsection{Examples of astronomical applications}

The impacts of trans-relativistic cameras on astronomy should be multi-fold and may not be fully appreciated until they become available. In the following, we discuss several examples that these cameras may make important contributions to astrophysics.

\subsubsection{Reionization history of the universe}
The first example is to study the reionization history of the universe \citep{loeb01,fan06}. The hot Big Bang theory predicts that the universe became neutralized around $z \sim 1100$, when electrons and protons recombined to produce neutral hydrogen. This is the epoch of the cosmic microwave background radiation. The universe was later reionized by the first objects (first stars, galaxies, and quasars) in the universe that shine in UV and X-rays (above 13.6 eV). The epoch between recombination and reionization is sometimes called the ``cosmic dark ages''. Quasar observations suggest that reionization was nearly complete around $z\sim 6$ \citep{fan06}. However, the exact detail of the re-ionization history is not known. Competing models predict distinct neutral fraction as a function of redshift (e.g. \citealt{holder03} and references therein). In order to map the reionization history in detail, one needs to populate bright beacons in the redshift range from 6 to $\sim 20$, and measure the spectrum blue-ward of the redshifted Ly$\alpha$ line, i.e. $\lambda \leq (1+z)1215.67{\rm \AA}$ (or $h\nu \geq 10.2/(1+z)$ eV).  A mostly neutral intergalactic medium would essentially absorb emission in this regime, forming the so-called Gunn-Peterson trough \citep{gunn65}. The shape of the red damping wing carries the important information of the neutral fraction of the IGM in that epoch \citep{miralda98}. However, such studies are hindered by the fact that the feature is moved to progressively more infrared bands as $z$ increases, and that the sources are typically fainter at higher $z$ as well.

Sending high-$\beta$ probes towards a high-$z$ galaxies or quasars would make it more convenient to measure their redshifts and to study the detailed red damping wing of these objects. This is because the Gunn-Peterson trough is shifted to the bluer bands in the probe's frame and the source flux is enhanced. 

Table 1 presents the relevant parameters for relativistic astronomy for different $\beta$ values. The parameters include Lorentz factor $\Gamma$, maximum Doppler factor ${\cal D}_{\rm max}$, its third power ${\cal D}_{\rm max}^3$,  and the relevant redshift, $z_{\rm Ly \alpha} (\lambda)$, for the Ly$\alpha$ wavelength (1215.67 \AA) to be at a particular wavelength $\lambda$ in Frame $K'$. For example, with $\beta \sim 0.3$, $z \sim 10$ can be probed with a 1$\mu$m camera, and the source is brighter by $\sim 1$ magnitude. If $\beta \sim 0.6$ is achieved, $z \sim 10$ can be even probed with an R-band ($\lambda \sim 0.658 \mu$m) camera, and the source is brighter by more than 2 magnitudes. Since an optical grism is much easier to build, one can use relativistic cameras with different termination speeds to probe a range of redshifts to most thoroughly study the reionization history of the universe.

\begin{table}
%\tablewidth{500pt}
\caption{Relevant parameters for relativistic astronomy for different $\beta$ values.}
\centering
\begin{tabular}{ccccc}
\hline\hline
  $\beta$  & $\Gamma$ & ${\cal D}_{\rm max}$ & ${\cal D}_{\rm max}^3$ & $z_{\rm Ly \alpha} (\lambda)$   \\
\hline
   0 & 1 & 1 & 1 & 8.2259 $(\lambda/{\rm 1 \mu m}) - 1$ \\
   0.1 &  1.0050  &  1.1055  & 1.3512 & 9.0941 $(\lambda/{\rm 1 \mu m}) - 1$  \\
   0.2 &  1.0206  &  1.2247  & 1.8371 & 10.0746 $(\lambda/{\rm 1 \mu m}) - 1$ \\
   0.3 &  1.0483  &  1.3628  & 2.5309 & 11.2100 $(\lambda/{\rm 1 \mu m}) - 1$ \\
   0.4 &  1.0911  &  1.5275  & 3.5642 & 12.5653 $(\lambda/{\rm 1 \mu m}) - 1$ \\
   0.5 &  1.1547  &  1.7321  & 5.1962 & 14.2477 $(\lambda/{\rm 1 \mu m}) - 1$ \\
   0.6 &  1.2500  &  2.0000  & 8.0000 & 16.4518 $(\lambda/{\rm 1 \mu m}) - 1$ \\
   0.7 &  1.4003  &  2.3805  & 13.4894 & 19.5816 $(\lambda/{\rm 1 \mu m}) - 1$ \\
   0.8 &  1.6667  &  3.0000  & 27.0000 & 24.6777 $(\lambda/{\rm 1 \mu m}) - 1$ \\
   0.9 &  2.2942  &  4.3589  & 82.8191 & 35.8559 $(\lambda/{\rm 1 \mu m}) - 1$ \\
   0.95 & 3.2026 &  6.2450  & 243.555 & 51.3708 $(\lambda/{\rm 1 \mu m}) - 1$ \\
   0.99 & 7.0888 & 14.1067 & 2807.20 & 116.0408 $(\lambda/{\rm 1 \mu m}) - 1$ \\
\hline
\end{tabular}
%\tablecomments{The hydrogen column density of Milky Way is fixed at $1.0\times10^{\rm 21} {\rm cm}^{\rm -2}$. Optical extinction and neutral hydrogen absorbtion of soft X-rays of the GRB host galaxy are taken into account, but they are negligible.}
\label{tab-SED}
\end{table}

\subsubsection{Redshift desert}

The redshift interval $1.4 \lesssim z \lesssim 2.5$ has been described by some authors as the ``redshift desert'' due the lack of strong spectral lines in the optical band $(4300-9000) {\rm \AA}$. Since this redshift range coincides with the epoch of significant star formation, the lack of a large sample of galaxies in this redshift range hinders an unbiased mapping of the star formation history of the universe  (\citealt{steidel04} and references therein). Observations with trans-relativistic cameras can easily fill this gap. One does not need a very high $\beta$ in order to achieve this goal. For example, with $\beta = 0.2$, one has a Doppler factor range from ${\cal D}_{\rm min}=0.8165$ to ${\cal D}_{\rm max} = 1.2247$ (depending on the observational direction $\theta'$). This is already enough to remove the redshift desert. In particular, for the redshift range $(1.9, 2.5)$, one can use the $\theta'=0$ mode, so that the effective redshift range in frame $K'$ is changed to $1.2247 \times (2.9, 3.5) - 1 = (2.55, 3.29)$, which is outside the desert. Similarly, for the redshift range $(1.4, 1.9)$, one can use the $\theta'=\pi$ mode, so that the effective redshift range is $0.8165 \times (2.4, 2.9) - 1 = (0.96, 1.37)$, again outside the desert. 

\subsubsection{Gamma-ray bursts}
% \item 
Gamma-ray bursts (GRBs) are most luminous astrophysical objects in the universe. 
In the case of observing transient relativistic events such as GRBs, relativistic astronomy would allow the humankind to study a relativistically moving source by a relativistically moving observer for the first time, a scenario previously only imagined in a thought experiment. Catching the early laboratory-frame IR afterglow of a GRB using an optical camera on board a high-$\beta$ probe would help to identify very high-$z$ GRBs. Theoretical models suggest that GRBs might form as early as $z \sim 20$ when the first-generation stars die \citep[e.g.][]{meszaros10,toma11}. 
A $\beta \sim 0.7$ probe will probe $z \sim 18.6$ with a 1$\mu$m camera. A systematic study of these explosions in the very early in the universe will help to probe the deep dark ages of the universe \citep{tanvir09}.

A good fraction of GRBs (30\%-50\%) are ``optically-dark'' GRBs. Even though high-$z$ GRBs may comprise a portion, most of them may be embedded in dusty star forming regions, so that the optical emission is absorbed via dust extinction \citep[e.g.][]{perley09}. If a relativistic camera is launched after the trigger of a GRB and observation is carried out during the acceleration of the camera, given the same observational frequency, the camera would continuously observes a range of frequency towards the IR regime, which will catch the characteristic features of dust extinction. Combining afterglow modeling, one may also precisely map the extinction curve of the GRB host galaxy, which is currently poorly constrained \citep[e.g.][]{schady12}.

\subsubsection{Electromagnetic counterpart of gravitational waves}

The new era of multi-messenger astrophysics just arrived recently with the detection of the first double neutron star (NS-NS) merger system GW170817 and its associated GRB 170817A and multi-band electromagnetic counterpart \citep[e.g.][]{abbott17a,abbott17b}. One important phenomenon is the so-called ``kilonova'', a type of IR/optical transient arising from r-process of neutron-rich materials dynamically ejected during the merger \citep[]{li98,metzger10}. The kilonova associated with GW170817 appeared to have a ``red''-component and a ``blue''-component \citep[e.g.][]{villar17}, with the former likely associated with the high-opacity ejecta possibly involving heavy elements such as lanthanides. Understanding these events will greatly benefit from a careful study of the spectra in a broad range from IR to UV. In the future relativistic astronomy era, NS-NS and NS-BH mergers will be regularly detected by the next-generation gravitational wave (GW) detectors. Observations with the trans-relativistic cameras in the directions towards and away from the GW trigger direction, together with ground-based observations, will help to uncover the broad-band spectra of kilonovae, leading to an in-depth study of the NS-NS and NS-BH merger physics.

\subsection{Testing special relativity}

One can also use the observations of trans-relativistic cameras to test the principles of spectral relativity. There are two ways to do so.
%test the principle of special relativity using the measurements performed in flight of a trans-relativistic camera.  

1. As seen from Figure \ref{fig2} and discussed in Appendix B, the measurement of relative positions of three point sources in Frame $K'$ as compared with those measured in $K$ (on Earth) can uniquely solve the motion of the camera (the direction and the Lorentz factor). With this information, one can predict the positions of the 4th, 5th ... point sources in the sky in Frame $K'$. If within the field of view of the camera there are more than three sources, a comparison between the observed and predicted positions in Frame $K'$ offers a unique test of the effect of aberration of light in special relativity. So far, a direct test of aberration of light has been made via observing the parallaxes of distant stars (through the very non-relativistic motion of Earth orbiting the Sun) \citep[e.g.][]{hirshfeld01} or via Earth-based experiments to measure a small gravitational aberration of light \citep[e.g.][]{kopeikin07}. A trans-relativistic camera will open the window to test this effect in the relativistic regime. 

% \item %The above applications assume the validity of special relativity. 
2. A comparison of the observations of a same bright object in two different frames ($K$ and $K'$) at the same intrinsic frequency offers another way to test the principle of special relativity. In non-standard theories such as massive electrodynamics, the Doppler factor ${\cal D}$ may take a form that slightly deviates from the simplest form Eq.(\ref{eq:Doppler}). A tight upper limit on the deviation of the measured (specific) flux at frequency $\nu'={\cal D}\nu$ in Frame $K'$ and that at $\nu$ in Frame $K$ (properly correcting for the Doppler boosting effect) would give a tight constraint on the deviation of ${\cal D}$ from Eq.(\ref{eq:Doppler}), and hence, the violation of the principle of special relativity. No similar test has  been performed so far.

\section{Conclusions and Discussion}

If indeed a trans-relativistic camera can be launched in the near future as envisaged by the ambitious Starshot project, besides the exciting prospects of sending cameras directly to Alpha Centauri, one can observe the universe with these cameras in flight. We have shown that unique observations can be carried out thanks to several relativistic effects. In particular, due to Doppler blueshift and intensity boosting, one can use a camera sensitive to the optical band to study the near-IR bands. The light aberration effect also effectively increases the field of view of the camera since astronomical objects are packed in the direction of the camera motion, allowing a more efficient way of studying astronomical objects. These observations also offer unique ways to test the principles of special relativity. We have discussed several examples of astrophysics research directions in which relativistic astronomy can play an important role to advance the fields.

According to the ``Breakthrough Starshot'' website, technology is being developed such that in the near future launches of low weight cameras are possible with low cost. An ambitious goal of these launches is to send probes to Alpha Centauri. From the above discussion, one can see that another powerful application of the technology would be to launch these cameras in any direction to study astronomical objects as needed. Starshot may become an observatory to launch cameras in desired directions with desired Lorentz factors in order to carry out astronomical observations.

According to Table 1, the relativistic effect is mild at 20\% speed of light, which is the target speed of StarChips to visit Alpha Centauri. The maximum Doppler factor ${\cal D}_{\rm max}$ is $\sim 1.2247$. Nonetheless, observable relativistic effects would take place. One can already make interesting tests of special relativity through comparing the images and rest-frame fluxes of bright objects as observed in Frames $K$ and $K'$, respectively. If one drops the goal of reaching Alpha Centauri, cameras with even higher Doppler factors may be designed and launched. A Doppler factor of 2 and 3 (which gives a factor of 2 and 3 shift of the spectrum) is available at 60\% and 80\% speed of light, respectively. More interesting astronomical observations can be carried out at these speeds.

For the Starshot project to send cameras to nearby stellar systems, two main challenges include how to shield cosmic rays during the long journey to the destination and how to transmit image data back to Earth from a large distance. For the probes launched for astronomical observations, these two challenges are somewhat alleviated, since there is no need to operate the camera for a long period of time or for a very large distance. In fact, given the same emitting power from the probe, the transmission signal received from Earth by a $\beta \sim 0.8$ probe at a light-hour distance is much stronger than that by a $\beta \sim 0.2$ probe at a light-year distance. The main challenge, on the other hand, is how to achieve a higher $\beta$ than the nominal value 0.2 (e.g. $\beta \sim 0.6$ needed to study $z\sim 10$ universe at R-band). This requires a laser accelerator with even higher power. For the purpose of studying high-$z$ universe, another issue may be the limiting magnitude limited by the small size of the cameras (even with the proper Doppler boosting). Technology for building large-area, thin and light chips is encouraged. If the obstacles to launch these chips to trans-relativistic speed are overcome in the future, unprecedented information about our universe can be obtained in the era of relativistic astronomy.

\acknowledgments
We thank an anonymous referee for valuable suggestions. This work is partially supported by NASA through grant NNX15AK85G.

\software{Python, https://www.python.org, \\ 
matplotlib, https://matplotlib.org, \\
GIMP, https://www.gimp.org}

\appendix

\section{A. Doppler transformations}

The standard Doppler transformation relations include \citep{rybicki79,begelman84,zhang17}: 
%Notice that the expressions are again opposite to the more familiar expressions used to describe the relations between a relativistic moving source and an Earth observer.
\begin{eqnarray}
 d t' & = & {\cal D}^{-1} d t, \label{t-D} \nonumber \\
 \nu' & = & {\cal D} \nu, \label{nu-D}  \nonumber \\
 E' & = & {\cal D} E, \label{E-D}  \nonumber \\
 ds' & = & {\cal D} ds, \label{ds-D}  \nonumber \\
 dV' & = & {\cal D} dV, \label{dV-D}  \nonumber \\
 d\Omega' & = & {\cal D}^{-2} d \Omega, \label{Omega-D}  \nonumber \\
 I'_{\nu'}(\nu') & = & {\cal D}^3 I_\nu(\nu), \label{Inu-D}  \nonumber \\
 j'_{\nu'}(\nu') & = & {\cal D}^2 j_\nu (\nu), \label{jnu-D}  \nonumber \\
 \alpha'_{\nu'}(\nu') & = & {\cal D}^{-1} \alpha_\nu(\nu). \label{alphanu-D} 
\label{D's1}
\end{eqnarray}
Here $dt$ is time interval differential, $\nu$ is frequency, $E$ is energy, $ds$ is length differential at source, $dV$ is volume differential at source, $d\Omega$ is solid angle differential, $I_\nu(\nu)$ is specific intensity of the emitting source at frequency $\nu$, $j_\nu(\nu)$ is the specific emission coefficient of the source at frequency $\nu$, and $\alpha_\nu(\nu)$ is the specific absorption coefficient at frequency $\nu$. The primed and unprimed quantities are with respect to Frames $K$ and $K'$, respectively.

Astronomy measurements concern the specific luminosity $L_\nu(\nu)$ and flux $F_\nu(\nu)$. The transformations of these parameters between the two frames depend on the source properties.

We first consider an isotropic, point source. In Frame $K$, one has
\begin{equation}
 L_\nu(\nu) = \int\int j_\nu(\nu) d \Omega dV = 4\pi j_\nu V,
\end{equation}
where $V$ is the emitting volume of the source, which for an optically thin source (e.g. a quasar) is the entire volume whereas for an optically thick source (e.g. a star) is the volume in the optically-thin region (i.e. within the skin-depth of last Thomson scattering).

In Frame $K'$, the source moves relativistically. One can write
%\begin{equation}
\begin{eqnarray}
 \frac{d L_{\nu'}(\nu')}{d \Omega'}  & = & \int j'_{\nu'}(\nu') dV' 
      =  \int {\cal D}^2 j_{\nu}(\nu) {\cal D} dV 
      %\nonumber \\
%\nonumber \\
%       & = & \int {\cal D}^2 \frac{L'_{\nu'}(\nu')}{4\pi V'} d\Omega \int dV  \nonumber \\
        =  {\cal D}^3 \frac{d L_{\nu}(\nu)}{d \Omega}
=  {\cal D}^3 \frac{L_{\nu}(\nu)}{4\pi},
\label{eq:dLnu-dOmega}
\end{eqnarray}
%\end{equation}
where the isotropic condition in Frame $K$ has been applied for the last equality. 

For a point source, all the emitter materials are considered to move towards one direction in Frame $K'$ (no $\theta$-dependence of the source material in terms of motion). 
%Due to relativistic beaming, the emitter
%is no longer isotropic in the lab frame, and most emission is beamed in direction
%of motion within the $1/\gamma$ cone. The ``true'' source luminosity has to be
%calculated by integrating Eq.(\ref{eq:dLnu-dOmega}) over $d\Omega$. On the other hand, an observer
An observer mostly cares about the {\em isotropic equivalent specific luminosity}, i.e. the specific luminosity {\em assuming} that the source is isotropic in Frame $K'$.
For a point source, the isotropic specific luminosity is simply Eq.(\ref{eq:dLnu-dOmega})
multiplied by $\int d \Omega' = 4\pi$, so that
\begin{equation}
 L'_{\rm \nu',iso}(\nu') = {\cal D}^3 L_{\nu}(\nu).
\end{equation}
The isotropic total luminosity at the frequency $\nu'$ is
\begin{equation}
 L'_{\rm iso}(\nu') = \nu' L'_{\rm \nu',iso}(\nu') = {\cal D}^4 (\nu L_{\nu}(\nu)).
\end{equation}

For an extended source, one should consider different spatial elements with different angles with respect to the direction of motion and integrate Eq.(\ref{eq:dLnu-dOmega}) by considering the transformation of the solid angle ($d\Omega' = {\cal D}^{-2} d \Omega$). At any angle $\theta$ between the direction of motion and line-of-sight in Frame $K'$, the {\em specific luminosity of a unit emitting element} at a particular
frequency $\nu'$ reads
\begin{equation}
L'_{\nu'} (\nu')= {\cal D} L_{\nu}(\nu).
\label{eq:Lnu}
\end{equation}
%This can be also derived by 
%\begin{equation}
%L'_{\nu'} (\nu')=\frac{dE'}{dt' \ d\nu'} = \frac{{\cal D} dE} {{\cal D}^{-1} dt \ {\cal D} d\nu} = {\cal D} \frac{dE}{dt \ d\nu} = {\cal D}
%L_{\nu}(\nu).
%\end{equation} 
The luminosity of a unit emitting element at a particular frequency $\nu'$ reads
\begin{equation}
 L'(\nu') = \nu' L'_{\nu'}(\nu') = {\cal D}^2 (\nu L_{\nu}(\nu)) = {\cal D}^2 L({\nu}).
\end{equation}

When considering the observed specific flux and flux of cosmological objects, one should differentiate the emission frequencies at the source $\nu_s$ ($\nu'_s$) and the observed frequencies $\nu$ ($\nu'$), which are related through the cosmological expansion factor $(1+z)$. Specifically, one has $\nu = \nu_s/(1+z)$ and $\nu' = \nu'_s/(1+z)$.  
%Strictly speaking, the frequencies in the above expressions of the luminosities are all relevant to the source frame, i.e. the relevant frequencies should have the `s' subscript. In the following, we restore 
We restore these subscripts and still define the observed frequencies as $\nu$ and $\nu'$ in Frame $K$ and $K'$, respectively.
For the case of a point source, in Frame $K$, the observed specific flux and flux at frequency $\nu$ for an Earth observer can be written as
\begin{eqnarray}
 F_\nu (\nu) & =&  \frac{(1+z)L_{\rm \nu_s}(\nu_s)}{4\pi D_{\rm L}^2}, \\
 F (\nu) & =& \nu F_\nu (\nu) =  \frac{L (\nu_s)}{4\pi D_{\rm L}^2}.
%=  \frac{(1+z) {\cal D}^3  j'_{\nu'}(\nu') V'}{D_{\rm L}^2},
\label{eq:Fnu}
\end{eqnarray}
where $z$ and $D_{\rm L}$ are the redshift and luminosity distance of the object. 
%The $(1+z)$ factor has specifically inserted for completeness to describe emission from a cosmological emitting source. 
In Frame $K'$, the observed specific flux and flux at frequency $\nu' = {\cal D} \nu$ for an observer in the probe (the camera) would be
%therefore
\begin{eqnarray}
 F'_{\nu'} (\nu') & = & \frac{(1+z)L'_{\rm \nu'_s,iso}(\nu'_s)}{4\pi D_{\rm L}^2} 
  =  \frac{(1+z) {\cal D}^3  L_{\nu_s}(\nu_s)}{4\pi D_{\rm L}^2} = {\cal D}^3 F_\nu (\nu), \nonumber \\ 
 F' (\nu') & = & \frac{L'(\nu'_s)}{4\pi D_{\rm L}^2} =  \frac{ {\cal D}^4  L(\nu_s)}{4\pi D_{\rm L}^2} = {\cal D}^4 F (\nu). 
%=  \frac{(1+z) {\cal D}^3  j'_{\nu'}(\nu') V'}{D_{\rm L}^2},
\label{eq:Fnu'}
%\nonumber
\end{eqnarray}

Similarly, for an extended source, specific flux and flux for an emitting element are given by
\begin{eqnarray}
 F'_{\nu'}(\nu') & = & {\cal D} F_\nu (\nu), \nonumber \\
 F'(\nu') & = & {\cal D}^2 F(\nu).
 \label{eq:Fnu'-extended} 
\end{eqnarray}
%The total specific luminosity and luminosity of an extended source needs to be calculated
%by integrating over the entire equal-arrival-time surface, which we will introduce in
%Sect. \ref{sec:EATS} and \ref{sec:extended} below.
%\medskip

\section{B. Solving the parameters of motion with three bright point sources}

The parameters of the motion for a probe include the direction of motion and the dimensionless velocity $\beta$ (or Lorentz factor $\Gamma$). For a constant velocity probe, one needs to measure the sky positions of at lease three point sources in Frame $K'$ (e.g. $1'$, $2'$, $3'$ in Fig.2) and compare them against the sky positions of the same three objects in Frame $K$ (1, 2, 3) in order to determine the parameters of motion. 
%This can be proven as follows.
%Let us name the three sources as 1, 2, 3 and $1'$, $2'$, $3'$ in Frame $K$ and $K'$, respectively (Fig.2). 
Let us suppose that the direction of motion is at point 0 and $0'$, respectively, in the two frames. 
%For simplicity, we assume that $0$ and $0'$ fall within the area enclosed by $(1, 2, 3)$ and $(1', 2', 3')$, respectively.

\begin{figure*}
\plottwo{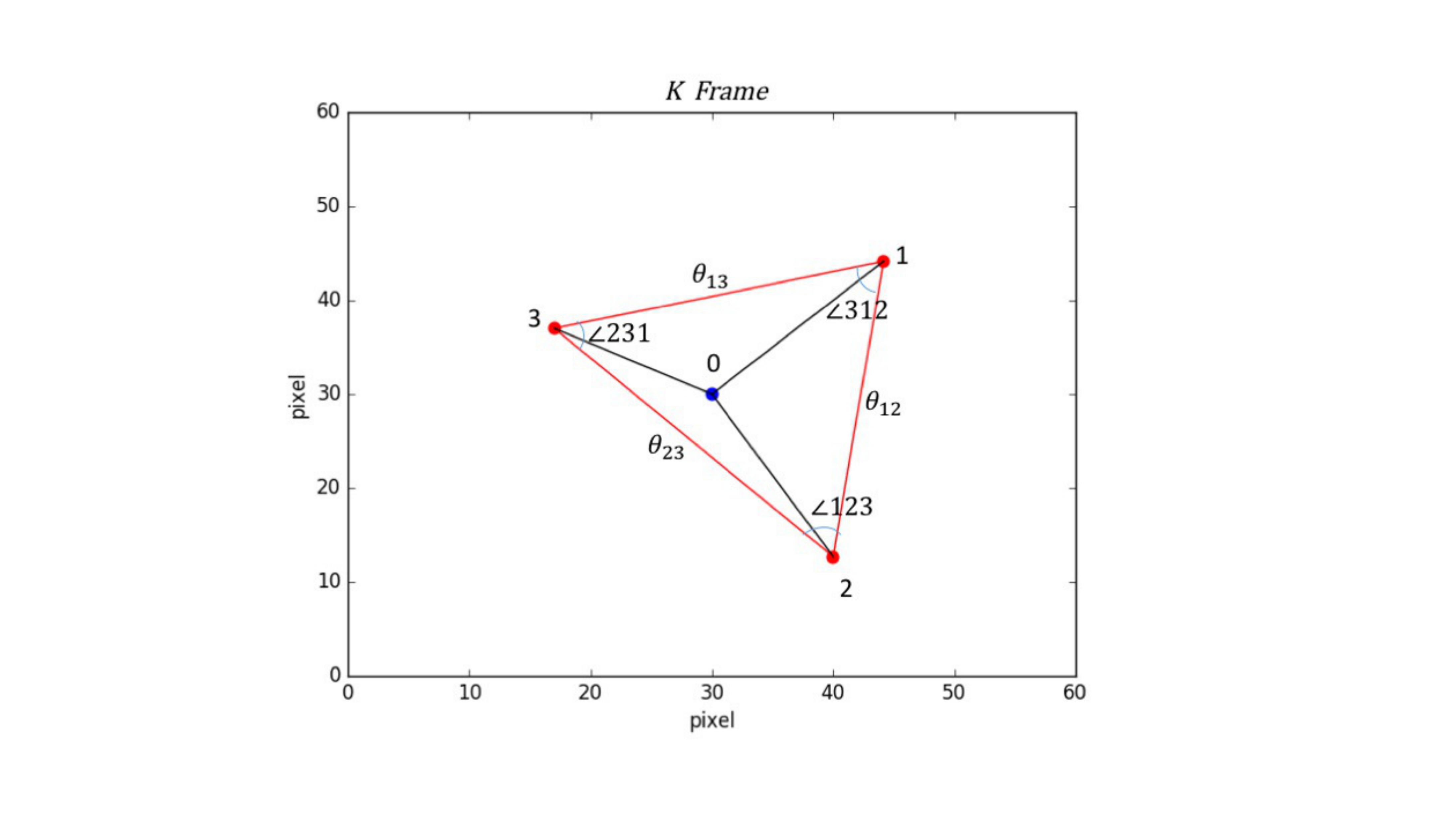}{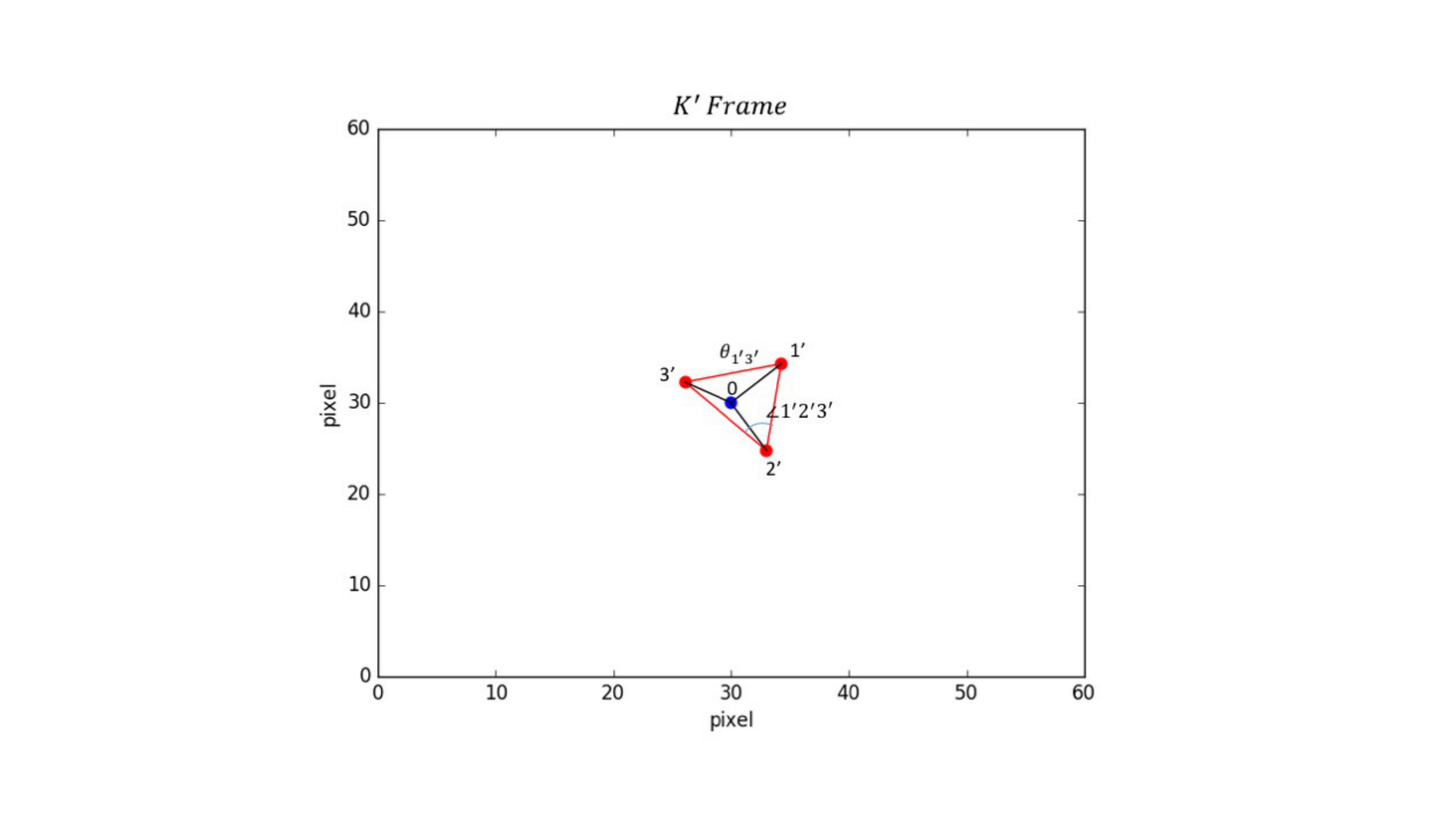}
%\begin{tabular}{c}
%\includegraphics[keepaspectratio,clip,width=0.5\textwidth]{f2a.pdf}%{f2a.eps} %{direction_K.jpg}  \\
%\includegraphics[keepaspectratio,clip,width=0.5\textwidth]{f2b.pdf}%{f2b.eps} %{direction_Kprime.jpg}\\
%\end{tabular}
\caption{The geometry to solve the motion of the probe. The three bright point sources are marked as 1, 2, 3, respectively, in Frame $K$ ({\it Upper}) and $1', 2', 3'$, respectively, in Frame $K'$ ({\it Lower}). The direction of motion is marked as 0 and $0'$, respectively, in Frame $K$ and $K'$. The relevant angular separations and opening angles are marked in Frame $K$ but not fully in Frame $K'$.
}
\label{fig2}
\end{figure*}

\begin{itemize}
\item The measured quantities include: the angular separations among the three points, i.e. ($\theta_{12}$, $\theta_{13}$, $\theta_{23}$) in $K$ and ($\theta_{1'2'}$, $\theta_{1'3'}$, $\theta_{2'3'}$) in $K'$ , and the respective angles among the three sources, i.e. ($\angle 123, \angle 231, \angle 312$) in $K$ and ($\angle 1'2'3', \angle 2'3'1', \angle 3'1'2'$) in $K'$.
\item There are 3+3+2+2+1=11 unknown quantities in order to solve the problem: the angular separations between the moving direction 0 (or $0'$) and the three sources, i.e. $(\theta_{01}, \theta_{02}, \theta_{03})$ in $K$ and $(\theta_{0'1'}, \theta_{0'2'}, \theta_{0'3'})$ in $K'$; the opening angles between 0 (or $0'$) and two of the three sources (that of the third can be uniquely determined if the first two are solved), e.g. $(\angle 012, \angle 023)$ in $K$ and $(\angle 0'1'2', \angle 0'2'3')$ in $K'$; and the dimensionless velocity $\beta$ (or Lorentz factor $\Gamma$). 
\item There are 4+4+3=11 equations making use of the above mentioned known and unknown parameters: In Frames $K$ and $K'$, there are four spherical triangles in each frame. Each triangle gives an independent equation according to the standard spherical geometry. Finally, there are three equations due to the light aberration effect, i.e. $\cos\theta'_{0i'} = (\cos \theta_{0i} + \beta) / (1 + \beta \cos\theta_{0i})$ ($i = 1,2,3$ and $i'=1',2',3'$).
\item As a result, the direction of motion and the dimensionless velocity of motion can be uniquely solved with the observations of three bright point sources in both frames. Such a solution is within the framework of special relativity.
% and the standard spherical geometry, the latter being under the assumption of a flat spacetime.
\item After solving the motion, one can add the observational data of the 4th, 5th ... celestial objects to the problem to test the validity of the above assumptions, e.g. predicting the positions in Frame $K'$ and use the observations to test the predictions. 
%Under the assumption of flat spacetime one may 
This leads to a test of the principle of special relativity. The precision of the test would increase with the increasing number of the sources observed. 
%Under the assumption of special relativity one may constrain the curvature of spacetime. Choosing sources at different distance bins (e.g. stars on our galaxy at a similar distance or quasars in a particular redshift bin), one may constrain the flatness of space at these different distance scales. 
\end{itemize}

\end{document}